\documentclass[12pt]{article}
\usepackage{graphicx}
\usepackage{psfrag}
\usepackage{hyperref}
\usepackage{adjustbox}
\usepackage{amssymb}
\usepackage[affil-it]{authblk}
\usepackage{natbib}
\usepackage{xcolor}
\usepackage{amsmath}
\def \IR{\hbox{{\rm I}\kern-.2em\hbox{{\rm R}}}}
\title{On the goodness-of-fit of generalized linear geostatistical models}
\author{Emanuele Giorgi}
\affil{ \small
Lancaster Medical School, Lancaster University, Lancaster, UK }

\begin{document}

\maketitle

\begin{abstract}
We propose a generalization of Zhang's coefficient of determination to generalized linear geostatistical models and illustrate its application to river-blindness mapping. The generalized coefficient of determination has a more intuitive interpretation than other measures of predictive performance and allows to assess the individual contribution of each explanatory variable and the random effects to spatial prediction. The developed methodology is also more widely applicable to any generalized linear mixed model.
\\
\\
{\bf Keywords:} coefficient of determination; generalized linear geostatistical models; goodness-of-fit. 
\end{abstract}

\section{Introduction}
\label{sec:introduction}
Generalized linear geostatistical models (GLGMs) are a class of mixed models where, conditional on a realisation of a Gaussian process $\mathcal{S} = \{S(x) : x \in A \subset \mathbb{R}^2\}$ in a study area $A$,  the outcome of interest $Y_{i}$, for $i=1,\dots,n$, follows a classical generalized linear model (GLM) \citep{GLM1989}.  Hence, the following properties hold.

\begin{itemize}
\item The $Y_{i}$, conditional on $\mathcal{S}$, are a set of mutually independent variables with mean 
$$E[Y_{i}| S(x_{i})]=m_{i}\mu_{i}=m_{i}g^{-1}(\eta_{i})$$ 
and variance 
$$\text{Var}[Y_{i}| S(x_{i})]=m_{i}V(\mu_{i}),$$ 
where: $m_{i}$ is an offset (e.g. number of trials for a Binomial response); $\eta_{i}$ is the linear predictor; $g(\cdot)$ is the link function; and $V(\cdot)$ the variance function.

\item $\eta_{i} = d(x_{i})^\top \beta + S(x_i)$ where $d(x_i)$ is a vector of explanatory variables associated with location $x_i$ and $\beta$ is a vector of regression coefficients.

\item The conditional distribution of $Y_{i}$ belongs to the exponential family.
\end{itemize}

In this technical note, we address the following question: how should we assess the contribution of the explanatory variables $d(x_{i})$ and of the random effects $S(x_{i})$ to our predictive inferences? \par 

To answer this question,  we propose a generalization of the coefficient of determination proposed by \citet{zhang2017} to GLGMs and show its application to a geostatistical data-set on river-blindness. For classical GLMs, Zhang's coefficient is defined as
\begin{equation}
\label{eq:r2_glm}
R^2_{GLM} = 1-\frac{\sum_{i=1}^n c_{V}(y_{i}, \hat{y}_{i}\{d(x_{i})\})}{\sum_{i=1}^n c_{V}(y_{i}, \hat{y}_0)},
\end{equation}
where: $\hat{y}_{i}\{d(x_{i})\}$ is the prediction for $Y_{i}$ based on $d(x_{i})$ by plugging-in the estimated regression coefficients via maximum likelihood; $\hat{y}_{0}$ is the prediction from a GLM with an intercept only; and 
$$
c_{V}(a,b) = \left\{\int_{a}^{b} \sqrt{1+[V'(u)]^2} \: du \right\}^2, \quad a, b \in \mathbb{R} 
$$
which measures the change in the variance function $V(\cdot)$ for a change in the mean from $a$ to $b$. When $Y_{i}$ is Gaussian and $g(\cdot)$ is the identity function, the numerator in \eqref{eq:r2_glm} reduces to the residual sum of squares, i.e. $c_{V}(y_{i}, \hat{y}_{i}\{d(x_{i})\}) = \sum_{i=1}^n (y_{i}-\hat{y}_{i}\{d(x_{i})\})^2$. \citet{zhang2017} also shows that \eqref{eq:r2_glm} does not overstate the proportion of explained variance by the explanatory variables compared to other generalizations of the coefficient of determination to GLMs that are based on the likelihood ratio \citep{maddala1983, cox1989, nagelkerke1991}. Furthermore, unlike the generalization by \citet{cameron1997} based on the Kullback-Leibler divergence, Zhang's coefficient of determination is also defined for quasi-models \citep{wedderburn1972} and, therefore, does not require the full specification of the likelihood function.

\section{A generalization of Zhang's coefficient of determination to GLGMs}
\label{sec:r2_glgm}
Our generalization of Zhang's coefficient of determination is based on the intuitive interpretation of random effects as accounting for the effect of unmeasured variables. \par

For simplicity, consider a GLM with two explanatory variables $D_{1}(x)$ and $D_{2}(x)$, hence
$$
\eta_{i} = \beta_{0} + \beta_{1}D_{1}(x_{i}) + \beta_{2}D_{2}(x_{i}), \quad  \text{for }i=1,\dots,n.
$$
Note that the two explanatory variables appear in the above equation in upper-case letters because we have not yet conditioned on them. Under such model, conditioning only on one of the two covariates might induce residual spatial correlation in the outcome  $Y_{i}$. Hence, if, for example, we condition on $D_{1}(x)=d_{1}(x)$, a natural model for the data would be a GLGM where $d_{1}(x)$ is used as an explanatory variable and $S(x)$ is used to account for the residual effect $\beta_{2}D_{2}(x)$. This argument can also be easily extended to any number of measured and unmeasured variables. \par
It follows that, conditionally on a realisation of $S^\top=(S(x_{1}),\ldots,S(x_{n}))$, a natural approach to quantify the total variation in $Y^\top=(Y_{1}, \ldots, Y_{n})$ is through
\begin{equation}
\label{eq:tot_variation}
\sum_{i=1}^n c_{V}(y_{i},\hat{y}\{d(x_{i}),S(x_{i})\}),
\end{equation}
where $\hat{y}\{d(x_{i}),S(x_{i})\}$ is the prediction for $Y_{i}$ based on the vector of explanatory variables $d(x_{i})$ and the realisation of $S(x_{i})$. Since $S$ is not observed, we can use its predictive distribution, defined as the distribution of $S$ conditional on $y^\top=(y_{1},\ldots,y_{n})$ and the covariates $d^\top=(d(x_{1}),\ldots,d(x_{n}))$ (henceforth $S|(y,d)$), to compute \eqref{eq:tot_variation}. More specifically, we average \eqref{eq:tot_variation} over the distribution of $S|(y,d)$, which leads to
\begin{eqnarray}
\label{eq:r2_glgm}
R^2_{GLGM} &=& 1-\frac{E_{S|(y,d)}[\sum_{i=1}^n c_{V}(y_{i},\hat{y}\{d(x_{i}),S(x_{i})\})]}{\sum_{i=1}^n c_{V}(y_{i}, \hat{y}_0)}
\end{eqnarray}
In the case of a linear geostatistical model, obtained by setting $m_i=1$, $g^{-1}(\eta_i)=\eta_i$ and $V(\mu_i)=\tau^2$ for all $i$, the expectation of \eqref{eq:tot_variation} reduces to 
$$
(y-D\beta)^\top(y-D\beta)+\xi^\top [\xi-2(y-D\beta)]+\text{tr}(\Omega),
$$
where: $D$ is a matrix of covariates;
$
\xi = \Sigma(\Sigma+I\tau^2)^{-1}(y-D\beta),
$
with $\Sigma$ and $I$ denoting the covariance matrix of the marginal distribution of $S$ and the identity matrix, respectively; and, finally,
$
\Omega = \Sigma-\Sigma(\Sigma+I\tau^2)^{-1}\Sigma.
$

For non-Gaussian responses, the expectation of \eqref{eq:tot_variation} is generally not available in closed form. We then propose to use a Monte Carlo Markov chain (MCMC) algorithm to simulate from $S | (y, d)$ and approximate \eqref{eq:r2_glgm} with
\begin{equation}
\label{eq:r2_glgm_mc}
R^2_{GLGM} \approx 1-\frac{\frac{1}{B}\sum_{j=1}^B\sum_{i=1}^n c_{V}(y_{i},\hat{y}\{d(x_{i}),s_{(j)}(x_{i})\})}{\sum_{i=1}^n c_{V}(y_{i},\hat{y}_{0})}  
\end{equation} 
where $s_{(j)}(x_{i})$ is the $j$-th out of $B$ Monte Carlo samples for the $i$-th component of $S|(y,d)$.\par

We can also define the coefficient of partial determination for the vector of explanatory variables $d$ given $S$ as
\begin{equation}
\label{eq:partial_r2_glgm}
\tilde{R}^2_{GLGM} = 1-\frac{E_{S|(y,d)}\left[\sum_{i=1}^n c_{V}(y_{i},\hat{y}\{d(x_{i}),S(x_{i})\}) \right]}{E_{S|(y,1)}\left[\sum_{i=1}^n c_{V}(y_{i},\hat{y}\{1,S(x_{i})\}) \right]},
\end{equation}
where $\hat{y}_{i}\{1,S(x_{i})\}$ is the prediction for $Y_{i}$ based on $S(x_{i})$ but excluding the explanatory variables $d(x_{i})$ from the model. We interpret \eqref{eq:partial_r2_glgm} as the fraction of explained variation in the response $Y$ by the explanatory variables $d$ but unexplained by the spatial random effects $S$. \par

In the next example, we compute \eqref{eq:r2_glgm} and \eqref{eq:partial_r2_glgm} by plugging-in the maximum likelihood estimates of the regression coefficients. These are obtained using the Monte Carlo likelihood method \citep{christensen2004} implemented in the R package PrevMap \citep{giorgi2017}. We simulate from $S|(y,d)$ using a Laplace sampling technique described in detail in Section 2.1 of \citet{giorgi2017}.

\section{Example: River-blindness mapping in Liberia}
\label{sec:rb_liberia}
River-blindness is an infectious disease caused by the parasite \textit{Onchocerca volvulus} and is transmitted by a black fly of the genus \textit{Simulium}. We analyse data from  90 communities in Liberia, where people were tested by palpation for the presence of skin nodules caused by the disease; for an Africa-wide analysis of these data, see \citet{zoure2014}. \par
\begin{figure}
\begin{center}
\includegraphics[scale=0.5]{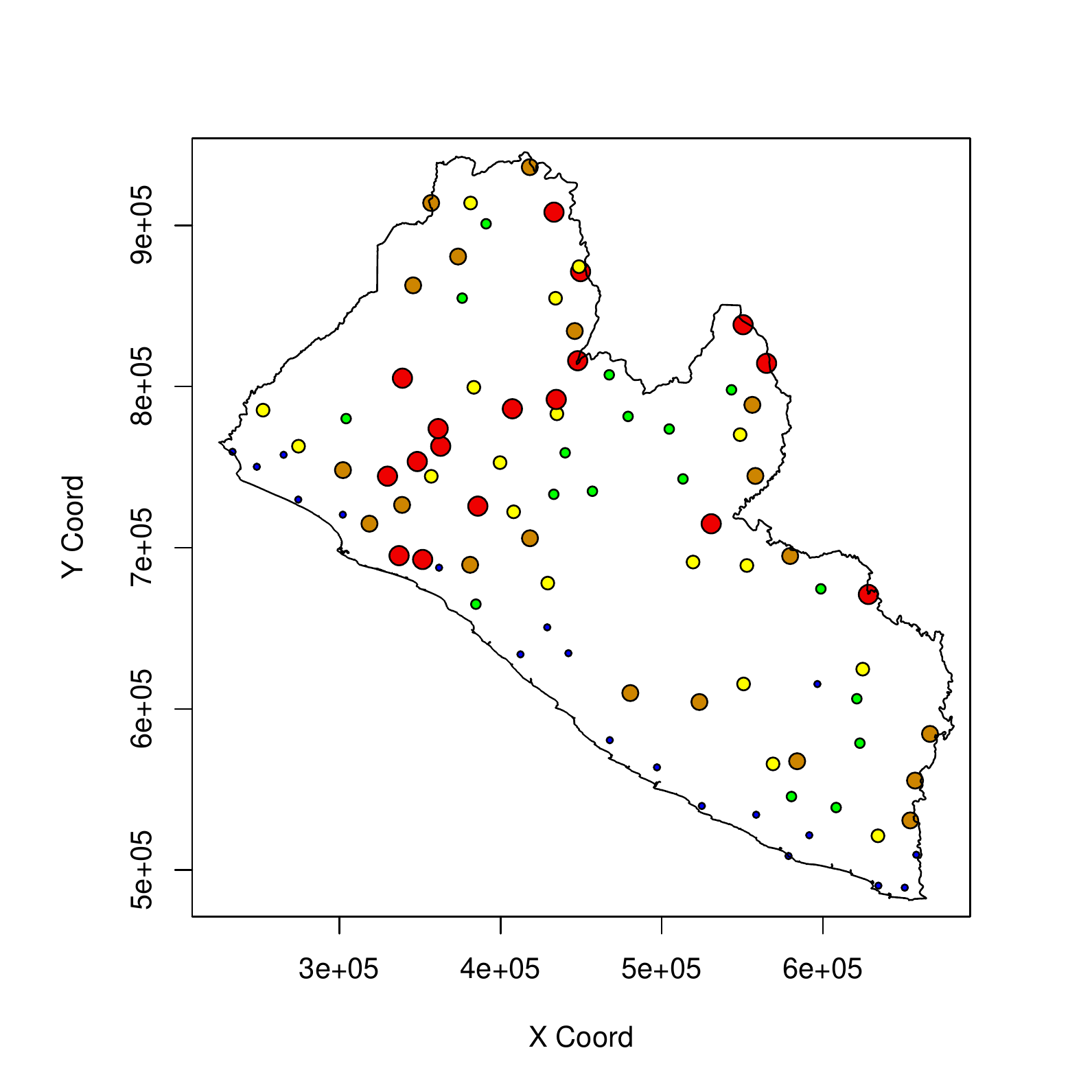}
\caption{Map of the empirical nodule prevalence. The radius of each point is proportional the quintile class within which the associated prevalence falls, with larger radiuses corresponding to higher quintile clasees. \label{fig:Liberia_rb}}
\end{center}
\end{figure}

Let $x_{i}$ be the location of the $i$-th sampled community, where $y_{i}$ out of $n_{i}$ randomly selected individuals tested positive. Our model for the data is a GLGM, where the $Y_{i}$ conditionally on $S(x_{i})$ are mutually independent Binomial variables with number of trials $n_{i}$ and probability of having skin nodules $p(x_{i})$, such that
\begin{equation}
\label{eq:lin_pred}
\log\left\{\frac{p(x_{i})}{1-p(x_{i})}\right\} = \beta_0 + \beta_1 x_{i,1}+ \beta_2 x_{i,2} + S(x_{i}),
\end{equation}
where $x_{i,1}$ and $x_{i,2}$ are the abscissa and ordinate components of the geographical location $x_{i}$. The reason for using a linear trend in $x_{i}$ is shown in Figure \ref{fig:Liberia_rb}, where the map of the empirical nodule prevalence shows an increase in the values as we move further from the coast in the north-east direction. Finally, we model $S(x)$ as a zero-mean Gaussian process with isotropic exponential covariance function having variance $\sigma^2$ and scale parameter $\phi$. \par 

\begin{figure}
\begin{center}
\includegraphics[scale=0.5]{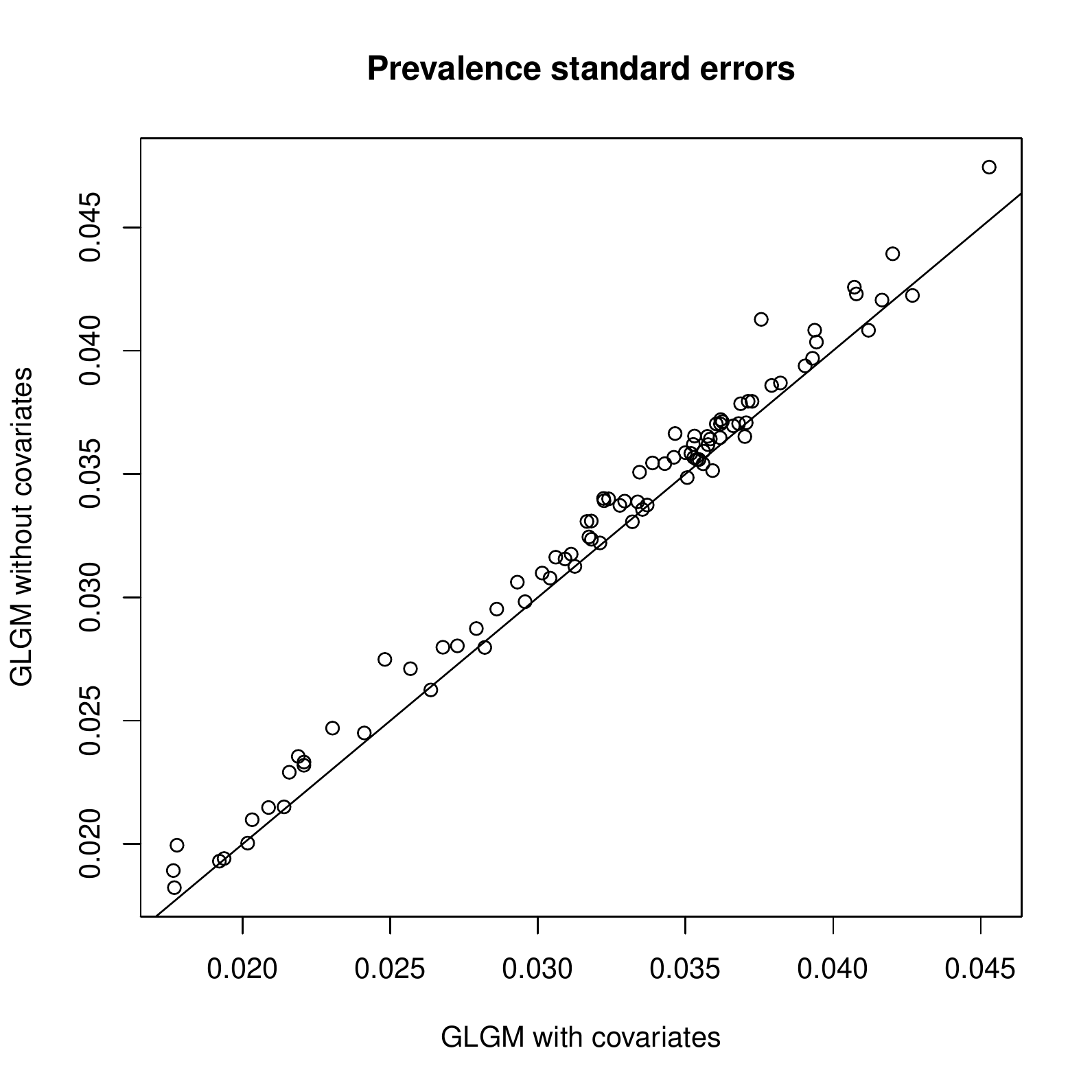}
\caption{Standard errors for the estimated nodule prevalence from a Binomial geostatistical model without covariates against one with covariates as defined in Section \ref{sec:rb_liberia}. The solid line is the identity line. \label{fig:std_errors_lib}}
\end{center}
\end{figure}

The maximum likelihood estimates  of the model parameters and their 95$\%$ confidence intervals are reported in Table \ref{tab:rb_liberia}. We observe that the use of the explanatory variables leads to a remarkable reduction in the values of the estimated $\sigma^2$ and $\phi$. The fitted GLGM explains about $59\%$ of the variation in nodule prevalence compared to $27\%$ from a classical GLM where $S(x)=0$ for all $x$. However, the small value of $1\%$ for the coefficient of partial determination, $\tilde{R}^2$, indicates that the point estimates from the GLGM with covariates, given by \eqref{eq:lin_pred}, are strongly similar to a model without covariates, where $\beta_{1}=\beta_{2}=0$. Nonetheless, Figure \ref{fig:std_errors_lib} shows that the standard errors for the estimated nodule prevalence (computed using Monte Carlo samples from $S|(y,d)$ while pugging-in the Monte Carlo maximum likelihood estimates) at the observed locations from the model with covariates are smaller almost everywhere than those from the model with only the intercept. More precisely, the largest relative reduction in the standard errors is of about $10\%$.

\begin{table}[ht]
\centering
\caption{Monte Carlo maximum likelihood estimates with associated 95$\%$ confidence intervals for the regression coefficients of the model with and without covariates defined in Section \ref{sec:rb_liberia}. \label{tab:rb_liberia}}
\begin{tabular}{rrcrc}
  \hline
  & \multicolumn{2}{c}{Without covariates} & \multicolumn{2}{c}{With covariates} \\
Term & Estimate & 95$\%$ CI & Estimate & 95$\%$ CI \\ 
  \hline
$\beta_{0}$ & -1.941 & (-3.312, -0.571) & -6.327 & (-9.126, -3.528) \\ 
$\beta_{1} \times 10^3$ &    &  & 2.761 & (0.223, 5.299) \\ 
$\beta_{2} \times 10^3$ &    &  & 4.784 & (2.208, 7.360) \\ 
$\sigma^2$ & 0.791 & (0.075, 8.295) & 0.145 & (0.055, 0.384) \\ 
$\phi$ & 395.050 & (32.608, 4786.143) & 68.526 & (20.438, 229.755) \\ 
   \hline
  \vspace{-0.2cm} \\
\multicolumn{5}{c}{$R^2_{GLM}= 27\%; R^2_{GLGM} = 59\%; \tilde{R}^2_{GLGM} = 1\%$} \\
  \vspace{-0.2cm} \\
   \hline
\end{tabular}
\end{table}

\section{Discussion}
\label{sec:discussion}
We have introduced a generalization of Zhang's coefficient of determination to quantify the proportion of explained variation in the outcome of interest by the covariates and/or the residual spatial random effects. This has a more intuitive interpretation than other measures of predictive performance, such as mean square errors, and also allows to quantify the individual contribution of each component of the linear predictor to spatial prediction. Although our focus was on geostatistical models, the developed methodology can be applied to any generalized linear mixed model. \par

Through an example on river-blindness mapping, we have quantified the impact of the adopted explanatory variables on the spatial estimates of prevalence. The proposed generalization of the coefficient of partial determination, $\tilde{R}^2_{GLGM}$, indicated that the impact of these on the point estimates of prevalence was negligible. We have also shown that the reduction in the standard errors, albeit small, was however more tangible than the change in the point estimates after adjusting for the north-east trend in disease prevalence. Hence, our recommendation is that $\tilde{R}^2_{GLGM}$ should not be used as a stand-alone tool but should be complemented with other measures that assess the impact on the accuracy of the spatial estimates. \par

Future research will aim to extend the methods of Section \ref{sec:r2_glgm} to point process models, including log-Gaussian Cox processes.

\section*{Acknowledgements}
Emanuele Giorgi holds an MRC fellowship in Biostatistics (MR/M015297/1).

\bibliographystyle{biometrika}
\bibliography{biblio}
\end{document}